# Precision-controlled ultrafast electron microscope platforms. A case study: Multiple-order coherent phonon dynamics in 1T-TaSe$_2$ probed at 50 femtosecond – 10 femtometer scales

Xiaoyi Sun[1], Joseph William[1], Sachin Sharma[1], Shriraj Kunjir[2], Dan Morris[2], Shen Zhao[2], and Chong-Yu Ruan[1*]

[1]Department of Physics and Astronomy, Michigan State University, East Lansing, MI, USA 48824

[2]Facility for Rare Isotope Beams, Michigan State University, East Lansing, MI, USA 48824

*Email correspondance : ruanc@msu.edu

**Abstract:**

We report the first detailed beam tests attesting the fundamental principle behind the development of high-current-efficiency ultrafast electron microscope systems where a radio-frequency cavity is incorporated as a condenser lens in the beam delivery system. To allow the experiment to be carried out with a sufficient resolution to probe the performance at the emittance floor, a new cascade loop RF controller system is developed to reduce the RF noise floor. Temporal resolution at 50 femtosecond in full-width-at-half-maximum and detection sensitivity better than 1% are demonstrated on exfoliated 1T-TaSe$_2$ layers where the multiple-order edge-mode coherent phonon excitation is employed as the standard candle to benchmark the performance. The high temporal resolution and the significant visibility to very low dynamical contrast in diffraction signals give strong support to the working principle of the high-brightness beam delivery via phase-space manipulation in the electron microscope system.

# I. Introduction

Transmission electron microscope (TEM) is a powerful materials characterization tool that offers nanoscale information on a wide range of materials and devices and the relevant technologies have been crucial for materials research[1]. As a structural probe, a main advantage with electron-based technology, instead of X-ray, lies in its flexible uses of field-based electron optics, providing multimodal operation, uniting crystallography, imaging, and spectroscopy with relative ease. However, TEM systems traditionally come with a very limited time resolution. Earlier efforts in upgrading the TEM with ultrashort time resolution encountered daunting roadblocks presented by the space-charge effect, leading to a significant trade-off between the resolution and the sensitivity[2]. Recently, a paradigm for obtaining high spatiotemporal resolution in ultrafast electron microscope (UEM) systems is pioneered by Zewail et al. [3, 4], based on lightly triggered photometers with few or less than one electron per pulse but operating at a high repetition rate (typically ranges from 1-100 MHz) [4]. This approach not only elegantly preserves the key features traditionally offered by the continuous wave (CW) TEM system, but also provides a baseline performance at sub-ps and sub-nm levels. Since the initial works, various types of UEM systems employing such one-electron-at-a-time paradigm have led significant impacts on many areas of research in nanophotonics[5-9], plasmonics[10, 11], and micro-analysis[12-15].

Meanwhile, boosting the current efficacy by operating at a higher bunch charge is necessary for a broad range of material research that typically does not have a short recovery time[1, 2]. In pursuing such high current efficiency systems[16-19], significant progresses have recently been made through constructive dialogs[1, 2, 20] between ultrafast electron diffraction, traditional electron microscopy, and the accelerator physics communities. Especially notable are the progresses in the ultrafast electron diffraction (UED) sector that yielded improved temporal resolution and/or boosted signal-to-noise ratio, exemplified in the construction of the 2$^{nd}$ generation DTEM at LLNL[21], incorporating radio-frequency (RF) cavity compression in keV UED beamlines[22-26], and advances in the MeV-UED systems with relativistic RF photoinjectors planned for the FEL injector at SLAC[27], BNL[28], DESY[29], and Shanghai Jiaotong University[30], among many others. A parallel development has been carried out at MSU where the RF compression element is used to condense the beam longitudinally in a high throughput arrangement to realize the ultrafast electron microscopes, targeting multimodalities[31, 32].

Here, we present the first detailed beam tests designed to benchmark the RF-beamline performance at the emittance floor, attesting the principles behind developing high current efficiency ultrafast electron microscope systems. The article is organized into principle and technical sections. In the principle sections, arguments derived from recent numerical studies are given for the laminar space-charge flow concept and the high-brightness beam delivery based on bunch phase-space manipulation. Derivations for the spatial and temporal resolutions from the RF-microscope both in imaging and diffraction modalities are introduced along with model simulations to set the performance baselines. The second part deals with the practical implementation in realizing the performance goals set by the beam phase-space areas, or so-called emittances. A key advance in stable control is made through introducing a new cascade RF feedback-control design, which effectively identifies and suppresses the phase noises to a level that when transcribed into the beam incoherence envelope is no more than the contribution expected from the bunch emittance floor. This implementation offers a means to test the performance at the fundamental stochastic limits in diffraction and imaging; the respective realm of operation set by the longitudinal and transverse emittances at bunch particle level of $10^5$ has the temporal and the spatial resolution under the scales of 50 fs and 10 nm. The beam tests are carried out by operating the microscope in the ultrafast electron crystallography modality on the exfoliated 1T-TaSe$_2$ thin film. With improved resolution and brightness, we target a more subtle scattering regime, involving multiple-order edge-mode phonons coupling to the intervalley scattering, which requires a gentler excitation and much less explored by a structural probe. We successfully record, at 5 kHz pump-probe repetition rate, the edge-mode coherent phonon excitation at 50 fs full-width-at-half-maximum (FWHM) resolution, 5 nm in coherence length, and sub-1% detection sensitivity. These levels of performance are highly commensurate with the theoretical predictions and thus validate the key concepts of implementing the high-currency ultrafast microscope systems.

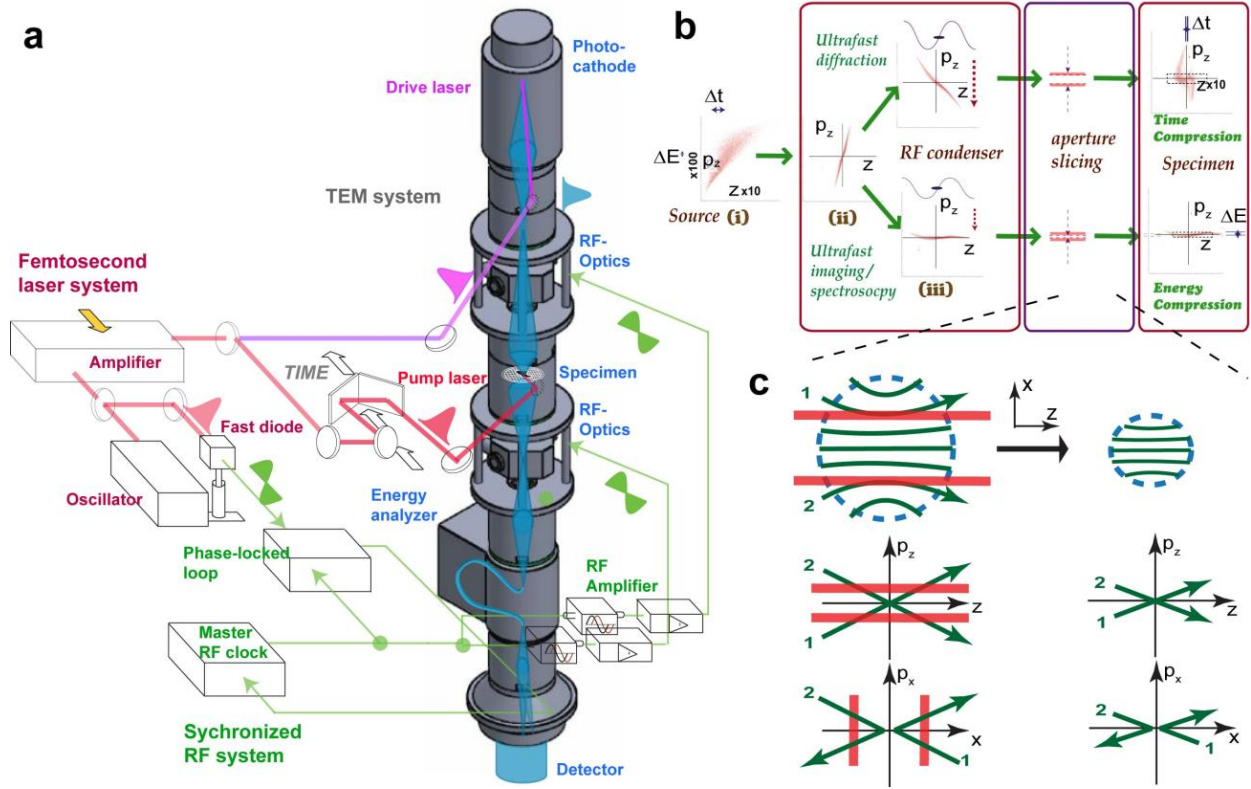

**Figure 1. The ultrafast electron microscope system with RF optics. a**. Schematic of the ultrafast microscope utilizing the RF cavities as electron lenses in a TEM column to temporally and spectrally focus the photo-electron pulses generated at photocathode. The fs laser system provides timing to pump pulse initiating material transformation as well as for seeding the RF signals sent to the RF cavity that delivers the short electron probe pulses. A phase-locked and control loop serves to ensure laser and RF are synchronized to each other. **b**. The phase-space manipulation to deliver time or energy compressed electron pulses. The key stages involve the phase-space chirp resetting following space-charge-led expansion, (i) to (iii), and condenser lens aperture slicing to adjust aspect ratios of the phase space structure while maintaining high core brightness. **c**. The mechanism of aperture slicing of high-intensity electron pulses that manifest in the laminar flow. The inter-dimensional effect is shown where due to non-mixing particle streams, exemplified with trajectories 1 and 2 in the peripheral areas, their removal with a condenser aperture causes tightening in both transverse and longitudinal phase space. *Adapted from Reference [19] with permission from RSC Publishing, Copyright 2023.*

## II. Performance at the phase-space emittance floor

Here we demonstrate by judiciously employing a beam phase-space manipulation protocol, one can approach a high current efficiency for running the UEM experiments without suffering major loss in spatial and temporal resolutions. Conventionally, the beam delivery in a transmission electron microscope builds on small area emitter where the electrons are extracted to form continuous single-particle stream; emission from narrow emitters with subsequent further condenser demagnetization provides the continuous wave TEM the capability to deliver low-emittance and high-coherence beam. However, in the consideration of photoemission that drives the ultrafast microscope with short electron bunches, the probe phase space is necessarily extended by the particle number in the bunch ($N_e$) and the beam brightness depends heavily on the packing density. Normally in a stochastic source where the individual particle trajectories may be considered as independent, the associated phase-space area (bunch emittance) would increase linearly with particle number; however, due to strong Coulombic interparticle interactions, the beam formation can develop fluid-like behavior. The best delivered beam is one where the bunch emittance grows sub-linearly with increasing $N_e$, leading to an increase of beam brightness - a regime which can be achieved via facilitating and maintaining laminar flow from the source to the specimen target. The condition to promote laminar space-charge flow under the flat cathode geometry has been investigated numerically based on the multi-level fast multiple method (ML-FMM)[33, 34]; here we will focus on the sequent beam delivery and the validation of the beamline performance leveraging the high brightness laminar flow. Specifically, to form coherent illumination from space-charge-dominated beam phase-space manipulation is required to reverse the velocity chirp effect derived from particle, which reduces the electron footprint but not the emittance size.

The phase-space manipulation is carried out through the joint application of magnetic (transverse) condenser and the new RF-cavity-based longitudinal condenser, and appropriate phase-space slicing considering the cross-dimensional effect. To illustrate the concept, the RF-enabled TEM optical beamline is schematically shown in Fig. 1a, where the high-intensity electron bunch is generated at a Pierce-gun photocathode via front-illumination by 266 nm drive laser pulses up to the virtual cathode limit. The flat cathode geometry and its proximity-coupling to a source condenser lens facilitate laminar flow emission just below the virtual cathode effect [34, 35]. The beam is fed into the condenser system of the UEM column, where, specifically, an RF cavity is inserted between magnetic lenses and serves as a new optical element to condense the bunch profile longitudinally. Meanwhile, the magnetic condenser and objective lenses demagnify the beam transversely. With the types of condensers acting together typically a pancake beam is formed through different lens settings. Just prior to illumination, a variable-size aperture array is employed to slice the emittance (and particle number) as required.

An important design principle for running the new UEM system, as discussed below, is to deliver the bunch in proper aspect ratio that best project the particle stream into the prioritized resolution window. This operates on the two extrema scenarios of focusing as illustrated in Fig. 1b, based on the longitudinal phase-space structure carried out by the RF optics – a similar picture equally applies to the bunch transverse focussing carried out by the magnetic condenser lenses. By running the RF-cavity field to reverse the longitudinal velocity chirp, the RF condenser refocuses the pulse to a smaller footprint along with positional or momentum direction to create a focused or temporally coherent pulse at the specimen plane; similar action from the magnetic condenser sets the illumination conditions as either spatially more focused (for imaging) or coherent (for diffraction). However, the applications of RF and magnetic condensers are not entirely independent from each other as when the beam is tightened longitudinally, it tends to expand transversely due to the internal space-charge forces.

The subtle cross-dimensional effect is a key element in considering the beam delivery in the new UEM system. We may contrast the different operations in terms of how the space-charge effect unfolds in a drastically different fashion in the CW and single-particle-stream UEM system. In these low-density beam systems, the presence of excessive independent particles tends to accelerate the stochastic growth in emittance causing degradation in performance. Meanwhile, the recent detailed fast multiple method simulation presented a different picture when propelling the photoemission into the high-density laminar flow regime[34, 36]. The basic physics behind this builds on the strong inter-particle correlation mediated through Coulombic interaction, which tends to develop bi-directional flows: the thermal particles with higher kinetic energy flow to the exterior whereas the laminar flow develops from condensing the low-entropy particles at the core region. Based on this bi-directional flow concept, a high-brightness regime for operating the UEM beamline can be conceived. This strategy targets the high-brightness core of the particle streams, as conceptually drawn in Fig. 1c, by applying an appropriately sized condenser aperture near the specimen. At this stage, the bi-directional flow has fully developed and the slicing along the transverse axis simultaneously removes particle streams that diverge longitudinally, more so in the circumference region than the core. This laminar-flow high-brightness strategy thus could result in a local boost in brightness for the beam delivered to the specimen. Such phase-space manipulation of high-intensity electron bunches becomes the foundation for operating the new type of RF-enabled UEM system with a high current efficiency.

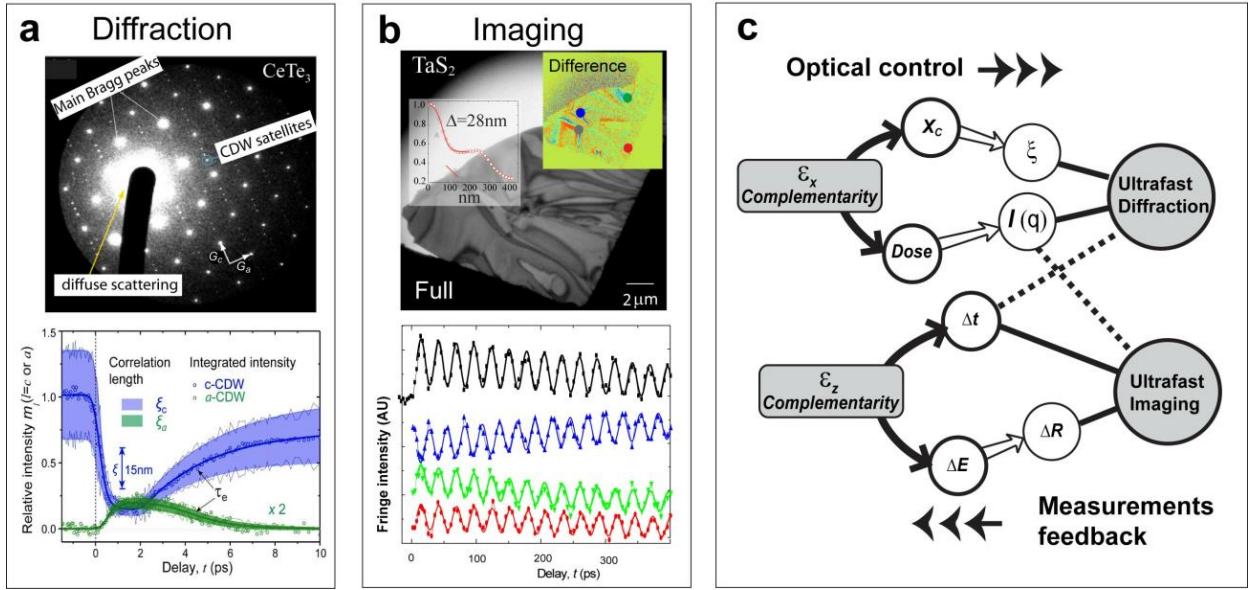

**Figure 2. Spatiotemporal resolutions from RF-optics-enabled UEM platform. a**. The ultrafast diffraction experiment carried out with $N_e$ of $10^6$ electrons on layered CeTe$_3$ thin film. The relative intensity $m_l$ gives the integrated intensity from structure factors derived from the CDW satellites. The correlation length $\xi$ and integrated intensity $I(q)$ derived from the structure factor at a given momentum gives an estimate for the bunch coherence length $X_c$ and the electron dose delivered. **b**. The ultrafast imaging experiment carried out with $N_e$ of $5\times10^5$ on layered 1T-TaS$_2$ thin film. The spatial ($\Delta R$) and temporal ($\Delta t$) resolution can be deduced from the intensity contrast and the dynamics of the bend fringe oscillations. **c**. The schematic intercorrelates the UED and UEM performance on the complementarity between different measurements. The scheme as detailed in the main text unites $X_c$, dose, $\Delta t$, $\Delta R$ over the bunch transverse ($\varepsilon_x$) and longitudinal ($\varepsilon_z$) emittances. The results set bounds on the emittances.

The bunch emittances estimated by the recent beam dynamics simulations under the laminar flow scenarios could be examined via methods targeting multi-modalities. Based on the theoretical principles, one can cross-correlate multiple figures of merit, including the coherence length ($X_c$), the dose, the imaging resolution ($\varDelta R$), and the temporal resolutions($\varDelta R$), relevant to the UED/UEM experiments based on the emittances of the bunch and the aberration coefficients of the electron optics. In another word, using the complementarily obtained performance metrics as the feedback, one could derive the bunch emittance; see Fig. 2c and discussion below. Such phase-space-centred beam optics arguments set a bound on the performance based on the stochastic emittance strongly influenced by the bunch charge. Following the scaling law established previously[34], it is predicted that, concerning the temporal resolution, sub-100 fs-level (in full-width half maximum, FWHM) RF focussing is achievable with $N_e$ up to $10^6$ (also shown in Ref. [37]); and accordingly, by slicing $N_e$ down to $10^5$, from the reduced emittances[19, 25] sub-30 fs-level temporal resolution and ≈1 nm-level spatial resolution could be obtained with the systems optimized for diffraction and imaging modalities. This level of performance has yet been fully demonstrated in the RF-optics-enabled UEM systems being implemented[31, 38]. The best recorded joint spatial-temporal resolution at MSU UEM systems is thus far beyond 100 fs-10 nm (as shown in Fig. 2) in spite the independently verified emittance figures support a higher level of performance. In the following, by examining how the bunch stochastic emittance and the instabilities in the RF optics control couple to the beam illumination incoherence envelope, we could better evaluate the impacts from both factors and attribute the noticeable discrepancy between the principal prediction and laboratory practice to either mismatched optical settings or appreciable beamline instabilities as compared to bunch stochastic emittance. Addressing the respective technical issues shall further allow the system performance to reach the stochastic floor set by the fundamental physics.

### III. Electron focussing in the UEM systems.

To understand the TEM optical manipulation with magnetic and RF condenser lens, we resort to the coupling between the phase space, considered as a stochastically filled area or emittance, to the optical transfer function through the objective aberration function[19]. To illustrate the theoretical principles, we first ignore the effects from a noisy RF source. We trace the effects from a finite-sized electron bunch phase space to the resolution function of the TEM. In this phase-space-based picture, the beam incoherent illumination causes information loss and degrades the resolution which can be modelled through the spatial and temporal incoherence envelopes[117]: $E_S(k)$ and $E_T(k)$, expressed in the Fourier spatial frequency $k$. As part of the (objective) lens transfer function[124], these incoherence envelopes restrict the information at high $k$, given by

$$t(k) = E_S(k)E_T(k)e^{i\chi(k)}, \qquad (1)$$

where λ is the electron de Broglie wavelength and $\chi(k)$ gives the phase factor of the distorted wavefront. The envelope functions take the exponential form:

$$E_S(k) = \exp\{-[\pi\sigma_s(C_S\lambda^2 k^3)]^2\}, \qquad (2)$$

$$E_T(k) = \exp\{-0.5[\pi\sigma_T(C_c\lambda k^2)]^2\}.$$

The direct impact from the incoherence phase-space envelope is seen in the coupling between exponent $\sigma_s$, the root-mean-square (RMS) of the beam convergence angle, and the spherical aberration coefficient $C_S$ that results in an expanded cone of incoherent illumination. Namely, with a large transverse emittance size $\varepsilon_x$, which propagates into a large transverse extremum size in terms of focusability (Fig. 1b), is expected to limit the spatial resolution ($R_0$) or the coherence length ($X_c$) from the bunch illumination. Meanwhile, concerning the temporal incoherence envelop where $\sigma_T$, defined by

$$\sigma_T = \left[\left(\frac{\sigma_E}{E_0}\right)^2 + \left(2\frac{\sigma_I}{I_0}\right)^2\right]^{1/2}, \qquad (3)$$

with $\sigma_E$ and $\sigma_I$ the RMS deviation of the beam energy ($E_0$) and objective current ($I_0$), couples to the lens chromatic aberration coefficient $C_c$. The expanded $\sigma_T$ limits the spatial resolution due to the cross-dimensional effect; namely beam temporal incoherence causes an imprecision in spatial focusing. It is easy to see the instability of the RF optics will couple to the lens optics through an increase in $\sigma_T$. Hence an imprecise RF lens not only causes a loss in temporal resolution, but also directly impacts the spatial resolution in an UEM. Here, based on the effect from the two incoherence envelops that we can evaluate, using the respective phase-space structures, the resolution loss is given by the complementarity (Fourier-pair) relation between the resolution function $R(r)$ and the transfer function $t(k)$[19]:

$$R(r) = FFT[\![t(k)]\!]. \qquad (4)$$

In Fig. 3, we present the results modelled for the UEM and UED modalities where the optical tunings of the transverse and longitudinal phase-space structures are carried out through the magnetic and RF condensers, measured in terms of the convergence half angle and the compressed pulse width $\Delta t$. For evaluating the effect on the focussing property, the results are given for different transverse and longitudinal emittances.

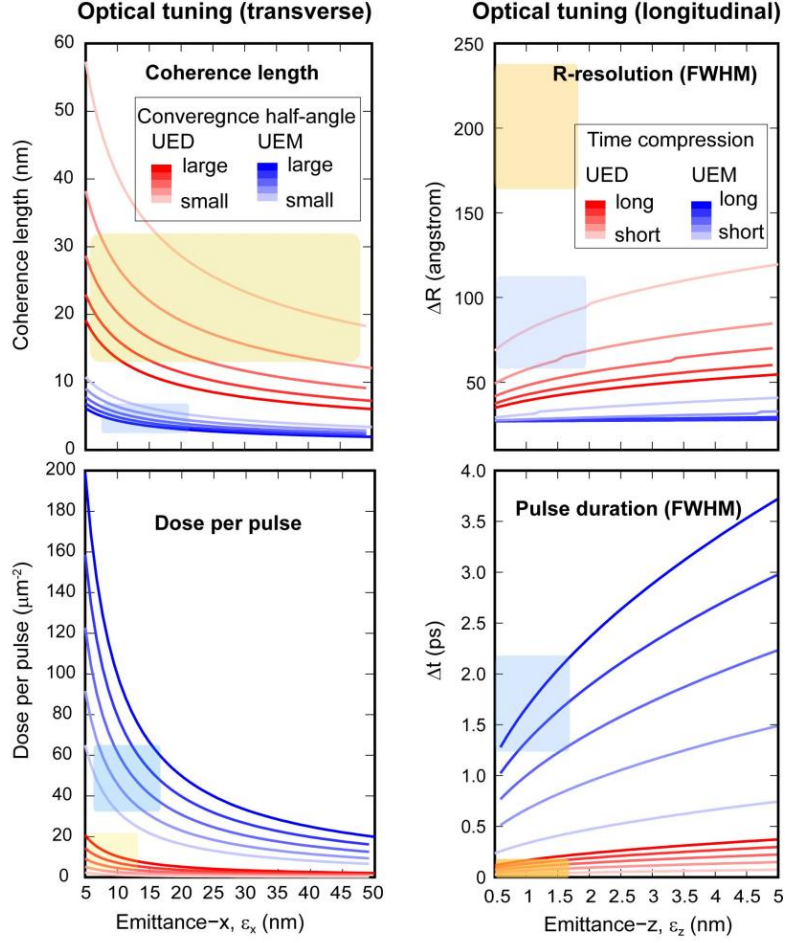

**Figure 3. UEM and UED performance from tuning the condenser system**. The left panels give the performance in the coherence length and electron dose under the tuning of the beam convergence half angle, a control parameter set by the magnetic condenser current and aperture size. The right panels give the performance in the imaging and time resolution under the tuning of the RF condenser. The model considers the beam energy of 60 keV with $N_e=10^5$. The UED and UEM categories are classified based on the tuning ranges. For UEM, the convergence half angle from 0.6 to 3.5 mrad is varied, subject to the demagnification of the beam by the condensers as a virtual source at the front focal distance of the objective prefield. The angle is subject to the value of $\varepsilon_x$, the condenser settings, and the prefield focal distance (1.4 mm). For UED, this angle is 0.1 to 1.1 mrad. Alternatively, in the tuning of the RF condenser, the longitudinal focussing $\Delta t$ is set for UEM from 250 to 750 fs (FWHM) at the low emittance, but nominally we increase this figure by the square root of $\varepsilon_z$ to simulate spatial resolution. For UED, the setting for $\Delta t$ is smaller, starting from 25 to 75 fs at low $\varepsilon_z$. The microscope resolution function from the derivation of $\Delta R$ is calculated using the lens aberration coefficients $C_S =1$mm and $C_C = 1.9$ mm.

Here, the categorization of modality is based on the starting values of the tunning parameters. As the tuning parameters are varied the performance in ultrafast diffraction and imaging may be prioritized; however, regimes where UED and UEM experiments are carried out at similar optical settings can be approached, namely a joint multi-modal scheme under the same beam illumination condition may be developed but with some compromises in resolutions from each channel. For operating in the imaging modality, optimized for spatial resolution ΔR, larger convergence half-angle is set to promote the dose over the coherence length, and longitudinally, a high temporal coherence is preferred over the temporal resolution to reach sub-10nm spatial resolution. In the case of ultrafast diffraction, the prioritization for higher $X_c$ and short pulse-width over the dose and direct-space resolution leads to a degradation in direct spatial resolution. The divergent trends from these simulations over the emittance provides the guidance, based off the delivery of performance in UED and UEM modalities, to set bounds on the size of emittance. The shaded areas represent the spans estimated from the experimental results operating the RF beamline under UED[39] (in red) and UEM[31, 32] (in blue) modalities; e.g., see Fig. 3. While the assignments give only the order-magnitude precision on the beam emittance, it is instructive to point out the assigned characteristic emittance is consistent with the source emittance given by the multi-level fast multipole method simulation[33, 34] optimized for the laminar flow regime.

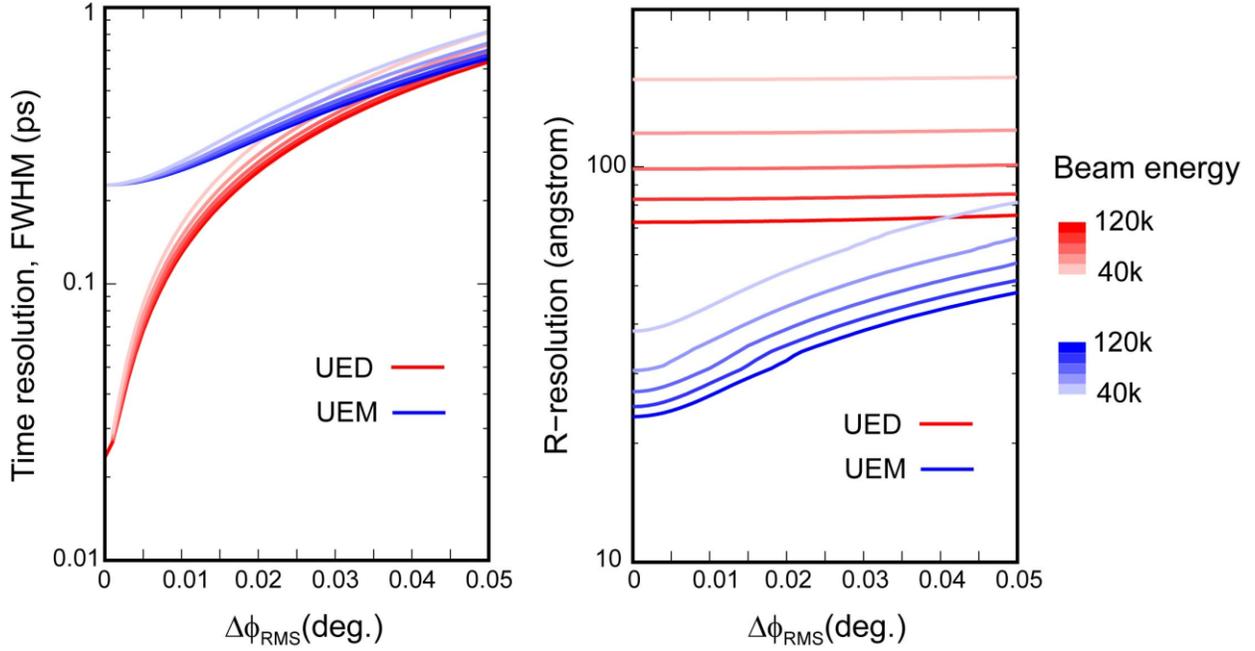

**Figure 4. The impacts from the RF instabilities on the temporal and spatial resolutions.** The left panel gives the temporal resolution for UEM and UED modalities modelled for the nominal values of $\varepsilon_x$ =10 nm and $\varepsilon_z$ =1 nm for $N_e=10^5$. The results are calculated for the beam energy varied from 40 to 120 keV. The right panel gives the corresponding changes in the imaging resolution. The different RF optical settings for UED and UEM reflects different phase-space aspect ratio prioritized for either imaging or ultrafast diffraction. For imaging the energy spread is 0.3 eV to allow for sub-10 nm imaging resolution, while for UED it increases to 250 eV to give a better temporal resolution set by the emittance floor.

We now trace the effects on the resolution from a noisy RF source feeding the RF-cavity longitudinal lens. To understand this effect on the performance, we relate the RF phase instability $\Delta\phi$ to an increase of temporal incoherence envelope given by the beam velocity spread[25]:

$$\frac{\Delta v_0}{v_0} = -\eta \frac{\gamma^2}{1+\gamma^2\beta^2} \Delta\phi, \qquad (5)$$

where $\gamma$, $\beta$ are the relativistic factors, and $\eta$ is the RF focussing parameter set by the cavity RF resonance field $\mathcal{E}(t) = \mathcal{E}_0 sin(2\pi f_0 t + \phi)$ at

$$\eta = \frac{e\mathcal{E}_0}{\gamma^3 \pi f_0 m v_0} sin\left(\frac{\pi f_0 d}{v_0}\right) cos\phi \qquad (6)$$

where $d$, $v_0$, and $\mathcal{E}_0$ are the bunch velocity, the gap size, and electric field amplitude of the cavity. Specifically, the values of $\eta$ to reach the two extrema shown in Fig. 1b is set by the focal distance $L_{lens-sample}$, with $\eta_E = \frac{a_0}{2\pi f_0}$ and $\eta_t = \frac{a_0 + v_0/L_{lens-sample}}{2\pi f_0}$ at bunch zero-crossing ($\phi = 0$ upon arrival at RF gap, see Fig. 1b), where $a_0$ is the bunch phase-space chirp, respectively for spectral and temporal compression. The corresponding arrival time jitter influencing the deliverable temporal resolution is given by

$$\Delta t_{arr} = \frac{\eta L_{lens-specimen} \gamma^2}{v_0(1+\gamma^2\beta^2)} \Delta\phi. \qquad (7)$$

Based on nominal emittance deduced for $N_e=10^5$ (Fig. 3), and the RF lens settings where $f_0 =1.013$ GHz, $d=2.1$ cm, we calculate the incoherence envelops following Eqns. (5)-(7) at the respective extrema for the UED and UEM experiments. The additional temporal incoherence from the RF cavity $\sigma_\phi$ is convoluted into the overall incoherence width by $\sigma_T = \sqrt{\sigma_\phi^2 + \sigma_E^2}$, from which we determine the spatial resolution. The results calculated for different beam energies are presented in Fig. 4 where we find the impact on the temporal resolution is largest in the ultrafast diffraction. We notice based on the simulation it is possible to compress the bunch down to 22 fs (FWHM) which only weakly depends on the beam energy. This weak energy dependence can be seen in the time-to-phase ratio, $k_{t\phi} = \frac{\eta L_{lens-sample}\gamma^2}{v_0(1+\gamma^2\beta^2)}$. With $L_{lens-specimen} \approx 0.4$ m in our setups, $k_{t\phi}$ of 4 to 4.5 is estimated for beam energy from 40 keV to 100 keV. Meanwhile, the impact on the ultrafast imaging lies in the spatial resolution; see right panel of Fig. 4. Irrespectively, the high sensitivity to the RF noise at the fs-nm scale measurements is quite noticeable, which highlight the importance for further improving the phase stability of the RF system. For testing the emittance floor based on $k_{t\phi}$ given in both scenarios, an RMS noise close to the level of 0.005° would be required.

### IV.     RF noise suppression with the cascade PID control system

While the core optical technologies combining magnetic and RF condense lenses promise to deliver a bright beam with sub-100 fs temporal resolution (*e.g.*, the simulation presented in Fig. 4 promises 22 fs FWHM at the specimen plane), translating such short bunch delivery to the performance requires a high stability and precision on the control system. For the UEM systems, the synchronization between the laser pump and the RF system seeding the RF cavity is achieved through a phased locked circuit. Given schematically in Fig. 1a, the circuit consists of a fast-rectifying photodiode that converts the light pulse signal from the laser oscillator into a frequency comb up to multi–GHz. The resonance frequency $f_0$ of the RF-cavity optics is locked on to the frequency comb at an integer harmonic of 1.013GHz. As the signals running from the laser frontend to the UEM cavity port is over an extensive distance, the success of the RF phased-locked circuit depends on identifying the spurious noise sources and their removal, which often involves proportional–integral–derivative (PID) control loops over the appropriate timescales of the experiments.

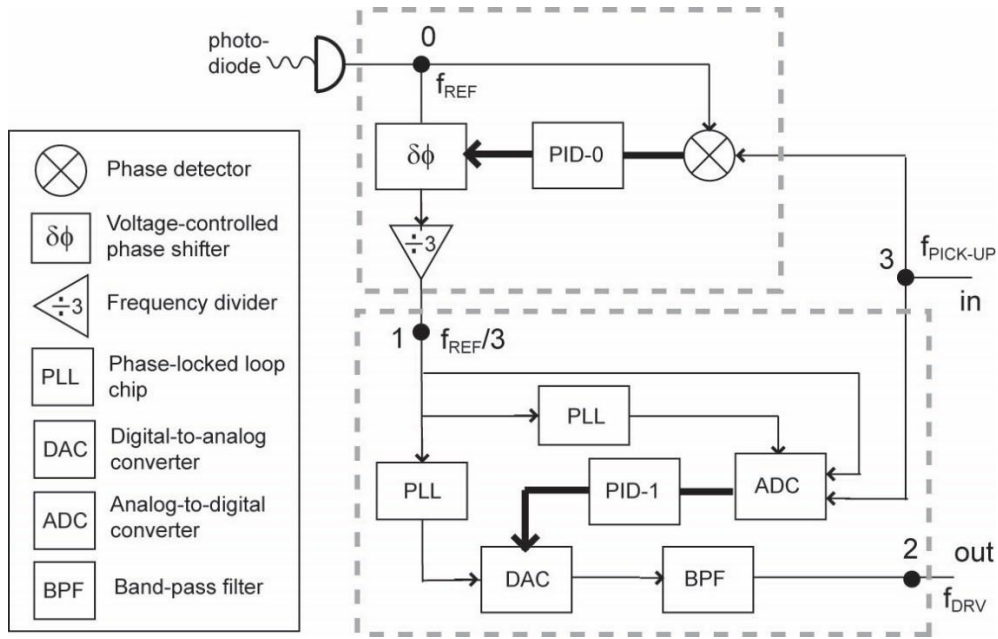

**Figure 5. Circuit schematics for the two-level PID system for RF phase control and noise suppression.**

Several effective strategies based on feedback-control have been reported in the literatures for running the RF systems used in the keV-scale UED[40, 41] and UEC/UEM[25, 39] experiments. Nominally, these systems have achieved a temporal resolution from 80 fs for the relativistic beams[30, 42, 43] to close to 100 fs for sub-relativistic systems[22-26], but the actual performance likely deviates from system to system due to varying beamline designs and the specific types of noise sources present in these locations. Here, we will focus on the development of a new PID scheme for the active noise reduction to reach the level of 0.005° RMS to provide sufficient precision to deliver the bunch focussing down to the emittance floor. The new scheme has a two-level cascade design as schematically drawn in Fig. 5.

Specially, a higher frequency PID-1 inner loop with a digital phase-locked loop (PLL) is proximity-coupled to the UEM column. It deals with the local phase noise $\Delta\phi_{13}$ measured between the signal fed into the low-level RF (LLRF) controller and the signal at the cavity pickup port. This high-speed LLRF controller is designed and fabricated by the RF group at Facility for Rare Isotope Beams (FRIB) to support the new 644 MHz superconducting cavity for future FRIB400 upgrade and 1.013 GHz room-temperature UEM cavity[44]. An important feature of this LLRF controller is the System-on-Chip (SoC) field programmable gate array (FPGA) design, fast data converters that allows direct sampling at 1.013 GHz and PLL chips with good phase noise performance up to 100 kHz that generates clocks for data converters and FPGA[45]. Thus, the LLRF controller can easily suppress any local phase noise within the cavity bandwidth ≈ 0.1 MHz.

Meanwhile, PID-0 is the master feedback-loop overseeing the entire length of the RF circuit. It compensates the phase difference, $\Delta\phi_{03}$, measured between the laser frontend and the RF pickup port; the output for the phase compensation is fed into the PID-1 subsystem. The PID-0 feedback-control employs a 4 MHz digitizer with averaging over 1k samples to reach sufficient phase detection precision to run the PID. The sub-millisecond detection time allows the system to detect noise spectrum beyond 1 kHz, yet the system operates at 10 Hz to give the overall phase stability it needs for longer term operation.

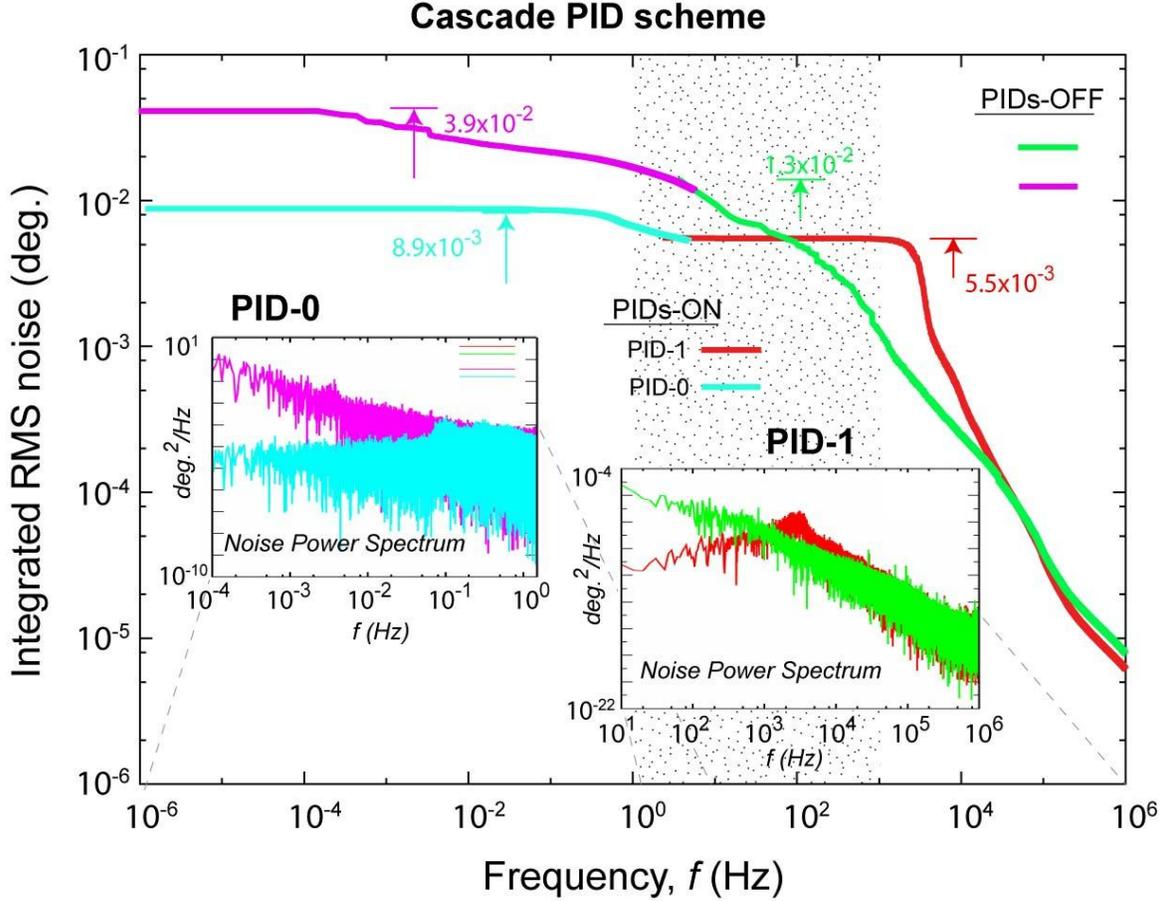

**Figure 6. RF noise characterizations before and after applying PID feedback controls.**

The nested feedback loops detect and compensate for the active noise sources over their own detectable spectral ranges. Importantly, the cascade loop design could synergistically reduce the noise floor as well as ensuring the longer-term stability through the existence of a shared responsibility spectrum region for joint optimization. This spectral region is allocated in the most active frequency domain from 10 Hz to 1 kHz. To inspect this effect, we first carry out the noise spectrum analyses at different levels of feedback controls by the two PID sub-systems. The results are given in Fig. 6, first shown with the scenario of open loops. In this free-running scenario, the intrinsic noise spectrum given by the PID-1 sub-loop carries an accumulated RMS noise that rises to 0.013° from 1 Hz to $10^6$ Hz (in green). Meanwhile, the phase noise detected in the master loop rises to 0.039° RMS from $10^{-6}$ Hz to 10 Hz (in pink). The corresponding noise power spectra are presented in the inset panels below, which show temporal correlations with the characteristics of a power-law distribution extended over several decades[46]. We then investigate the synergistic effect from running the two PID control loops simultaneously with the set parameters used in PID-0 optimized for the betterment of overall performance viewed from $\Delta\phi_{03}$, which necessarily encompasses the increased noise from the active PID-1 sub-loop for reducing the high frequency noises. At this stage of implementation, only the PID-0 parameters are tuned.

By activating both PID loops, the noise levels drop significantly. For the PID-1 subsystem, the integrated noise reaches the 0.0055° noise floor. In the PID-0 master loop, noises are added in the range above 10 kHz, reaching an overall integrated RMS noise of 0.0089°. By comparing the corresponding noise power spectrum, shown in the sub-panels below, it is evident that the improvements are achieved mainly by suppressing the diverging spectral power at lower frequencies; respectively, the feedback action defines a corner frequency $f_c$ of 0.5 Hz and 3 kHz for the effective noise suppression in the two loops. By setting the goal for reducing the integrated noise, the noise below $f_c$ is markedly reduced; however, a slight increase in noise power above $f_c$ may be observed as the result of active intervention.

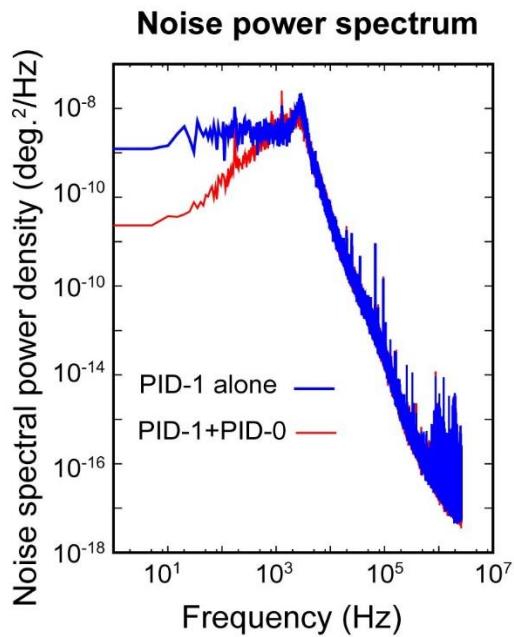

**Figure 7. Interactions between the two PID control loops in the 10 Hz to 1 kHz range.**

To study the interaction between the two control loops, we examine the time traces over different levels of PID controls. The blue curve in Fig. 7 shows the noise power spectrum from independently running the high-frequency feedback loop, where one can see the noises spectrum levels off at around 1 kHz. However, by turning on the PID-0 feedback control with the goal of reducing the overall integrated noise, the noise spectrum seen at frequency below 1 kHz drops markedly. We view this synergistic effect also from the perspective of PID-0 loop. The results are given at different levels of feedback controls, first presented in the time traces in Fig. 8a over the span of 2000 seconds. Here, as the reference the top trace (in red) shows the free-running scenario. The trace (in pink) below gives the results from activating just the PID-1 feedback control. The noise level is reduced but not significantly. This surprising result is understood by closer examination of the noise fluctuations (inset panel) where a visible telegraph-like noise with switching behaviour is shown. The switching frequency of this noise is below 0.1 Hz, thus undetected at PID-1 level. Locally within each plateau, the noise level is significantly lower than the free running case. From the histogram analysis, these noise features give bi-modal distribution and carry a significant part of the integrated noise; see Fig. 8b. We also examine the case with activating just PID-0; see Fig. 8a in green. In this case, the accumulated RMS noise reaches a level close to 0.01°, better than the previous scenarios; see Fig. 8c. But the local noise is higher than the case with just turning on the PID-1 feedback control albeit no signature of the telegraph noise can be traced here. We thus can conclude the presence of the telegraph noise to be a spill-over effect from active intervention in the higher frequencies – one that turning on PID-0 feedback control can help remediate. Indeed, by activating the two feedback loops simultaneously, both the local and integrated global noise are reduced.

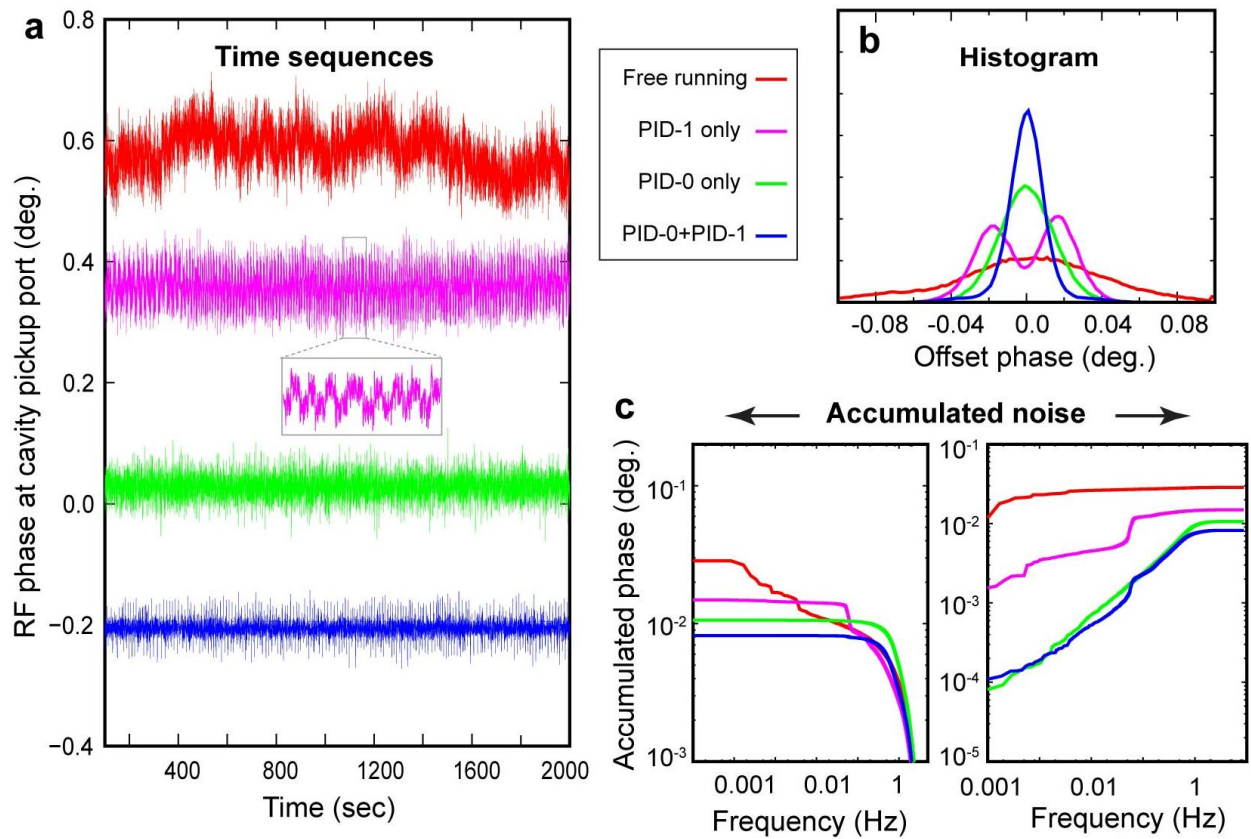

**Figure 8. Noise characterization under different PID scenarios**. **a**. The noise time sequences obtained by PID-0 phase detector with increasing levels of PID control; see the color-coded figure legend for the different PID settings.

A pronounced source of noises is given at the level of PID-1 feedback control, where the uncompensated noises from PID-1 subsystem overflow into the master loop in telegraph noise-like steps at a relatively low frequency, ≈0.06 Hz; see the inset. **b**. The phase noise histograms obtained from corresponding time sequences in panel a. **c**. The accumulated RMS noises derived from the time sequences. The left sub-panel shows the results integrated from the high-frequency spectrum; whereas the right sub-panel gives the results integrated from the low-frequency.

We further consider the ramification of the noise suppression protocols on the actual experimental implementation where ultrafast diffraction or imaging carried the information frame-by-frame over the acquisition time of seconds. From this perspective, we examine the RF phase independently monitored at the UEM column while running the experiments shown in Fig. 9. By comparing the results between just activating PID-1 and joint feedback control (PID-0+PID-1), one can judge the main improvement made in integrated RMS phase noise comes from reducing the active period of the telegraph noise, and upon averaging their contribution is minimized to a small pedestal region in the histogram representing the resolution function in the temporal response. Clearly, based on the Gaussian sigma value extracted from the histogram, the precision of the RF system has reached within 0.01° RMS. At this low level of jitters, it is essential to test the sampling error; currently to be sensitive to noise from the higher frequency region for joint feedback control the phase detector integration time is at 60 µs — which clearly is able to pick up the sharp rise of the telegraph noises (and hence correcting the effect in the feedback loop), but the short integration time may introduce a detector noise higher than the noise floor of the device. For this purpose, we deliberately raise the acquisition window to 350 µs; see Fig. 9 colored in green. At this acquisition time, the detector is still sensitive to kHz noises and can faithfully detect the residual noise spikes but long enough to reduce the sampling noises. The noise distribution as given in the corresponding histogram gives a much-reduced Gaussian width of 0.0053°. Given we are confident that the noise spectrum above 1 kHz is effectively suppressed by the PID-1 feedback control (Fig. 6), we believe the measurement here does reflect the noise floor from the joint PID control loop.

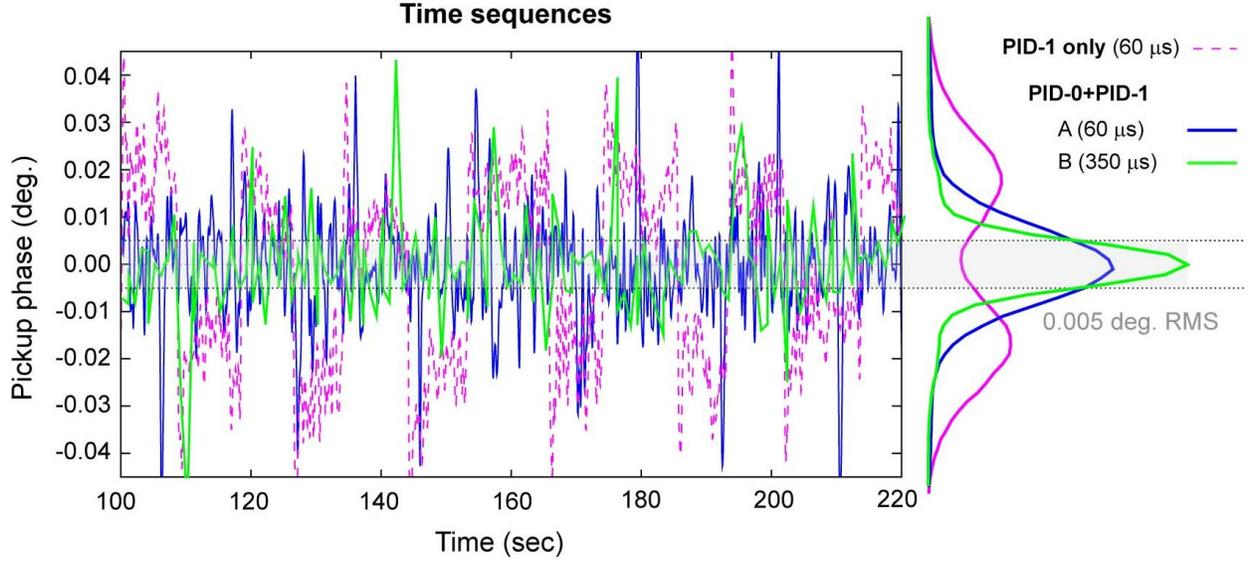

**Figure 9. RF phase stabilities tracked by phase detector at the UEM station.**

With real-time recording of the phase noise, one may adopt a strategy of rejecting the effects from spiky features by setting the acquisition primarily over the low noise regions. Keeping the acquisition time well below the timescale of the noise spikes allows such discrimination. Meanwhile, any residual noise distributed over the acquisition time can be effectively compensated through data processing using the recorded phase as the time stamp to shift the timing, following Eqn. (7). Judicious applications of these strategies could reduce the noise floor for the ultrafast measurements even further.

## V.     Implementations

The reduction of the RF system phase instabilities offers a platform to investigate the ultrafast structure dynamics down to the emittance floor; for particle numbers as few as $10^5$, one expects a new limit in the temporal resolution well below 100 fs to be investigated by the sensitive RF-optics-based UEM systems. Here, we explore this fundamental limit of probing with thin layered quantum materials, where ultrafast lattice responses well into the multi-THz regime could be well-isolated due to the reduced dimensionality. Especially, in the transition metal dichalcogenide (TMDC) system, it has been reported that the longer-lived valley polarization exists in mono- or few layers upon applying short pulses [47, 48]. The process directly involves coherent phonon emission to conserve the momentum from intervalley scattering. This coherent phonon generating mechanism is distinctly different from the more broadly investigated scenario where the lattice system collectively responds to the impulsively applied pump excitation. In this regime, the system perturbation rapidly redistributes the carriers and imposes a swift deformation force that activates structural modulation and shifts the electronic band-edge often detected through a change in the dielectric constant. Investigating such impulsively driven nonequilibrium band-lattice structure coupled dynamics within a typically very short time window presents a niche vista point to investigate the fundamental coupling hierarchy[49, 50]. With improved resolution and brightness, we aim at a more subtle quantum scattering regime, involving momentum-selective and layer-thickness-dependent behaviour, which requires a gentler probing and much less explored by a structural probe. Specifically, a recent study on $MoSe_2$ through the all-optical approach identified a non-sinusoidal waveform dominates the transient transmittance signals[51]. Guided by the density-functional theory calculation the specific dielectric response is mapped into a multiple-order longitudinal acoustic (LA) mode excitation that couples directly to the scattering from the Brillouin zone K to K' valley corners at 4.7 THz[51]. It would be instrumental to directly identify such valley-scattering-coupled phonon signatures with a structure-based probe. Moreover, achieving the visibility for isolating the specific nonlinear frequency response can serve as a standard candle for evaluating the temporal responses.

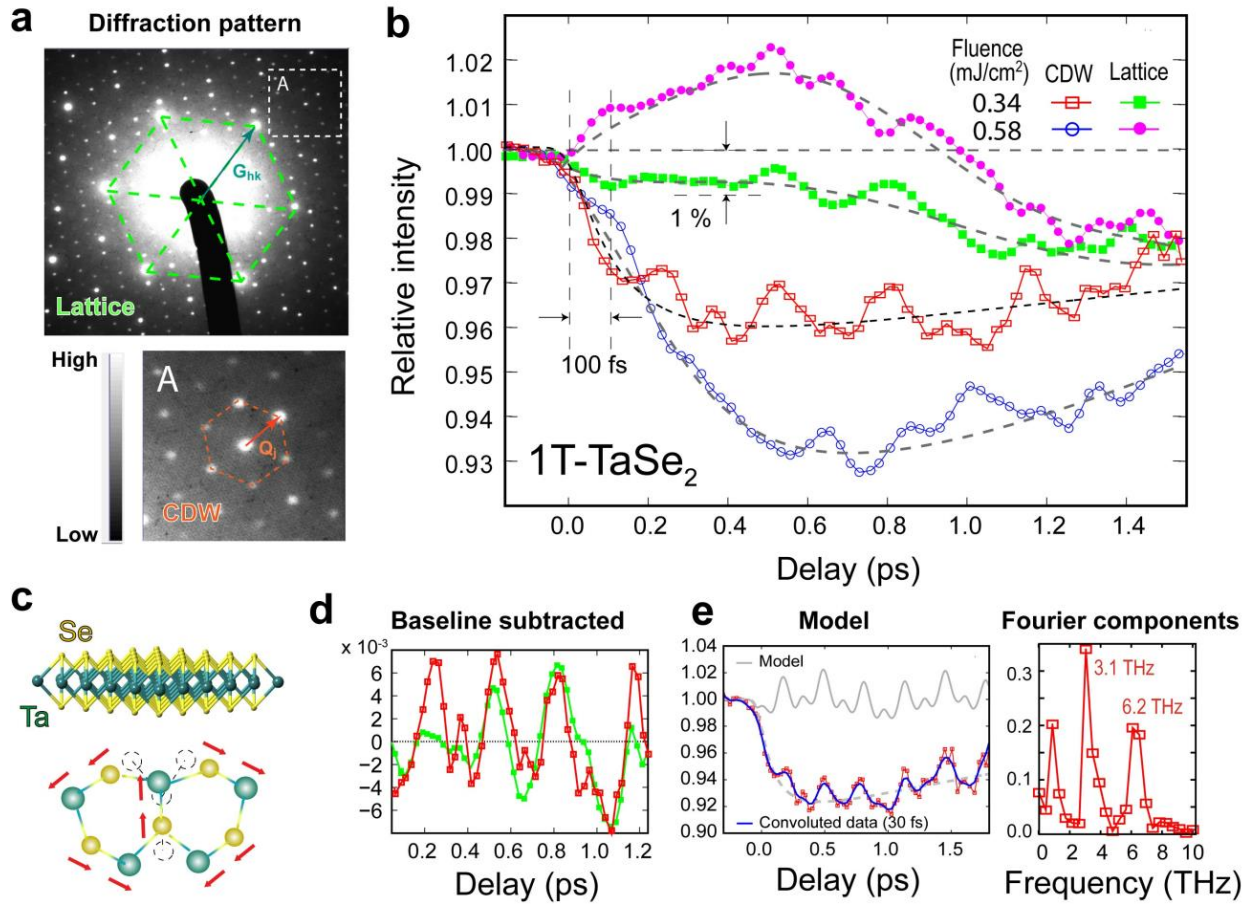

**Figure 10. Coherent phonon dynamics probed by ultrafast diffraction modality. a**. The diffraction pattern from the exfoliated 1T-TaSe$_2$ sample. The dashed lines outline the corresponding triangular lattice and CDW supercells periodicities. **b**. The intensity modulations registering the coherent phonon dynamics and the incoherent channels including the thermal phonon signatures and the density-wave state evolutions; see text for details. The coherent multiple-order nonlinear phonon excitation involving zone-edge (K) acoustic phonon in periodic high and low intensity modulation is visible in the low fluence data set. **c**. The modal structure of the K-edge phonon schematically reproduced from the previous study[51]. **d**. The baseline-subtracted transients from the main and superlattice structure factors indicating phase-locking. **e**. The Fourier analyses conducted to retrieve the coherent phonon frequencies.

We choose to investigate the TMDC family member 1T-TaSe$_2$, which has a similar in-layer triangular atomic lattice as MoSe$_2$. But the system also hosts many-body charge-density wave ground states[52, 53], accompanied by the opening of a bandgap at the relevant CDW wave-vector at the Fermi surface. The coexistence of the local extrema in the electronic band structure relevant to the valleytronic process and, in different parts of the Fermi surface, the CDW-mediated gap formation subjects this TMDC system to different types of tuning knobs for ultrafast controls. The experiments were conducted employing the generation-I UEM system at MSU. The 1T-TaSe$_2$ specimen was prepared through tape-exfoliation to ≈ 40 nm thickness over ≈ 25 μm in lateral size and pumped with 30 fs near-IR (800 nm) laser pulses at 45-degree incidence under the vacuum. We performed the pump-probe UED experiments with bunch charge limited to $10^5$ to confine the emittance in the nm scale to approach the required sub-100 fs temporal resolution. The scattering data integration was carried out at 5 kHz repetition rate over an 8-sec integration time to gather more than $10^7$ electrons in the relevant Bragg scattering. The overall sensitivity thus is expected to reach sub-1% over the intensity modulation (corresponding to < 1% or 10 femto-meter level bond-distance changes). In this feasibility study, we focus our work on illustrating the previously revealed edge-mode coherent phonon signatures involved in the quantum scattering scenario [51] that occurs well below the CDW phase transition threshold. We note that the coupled ultrafast structural and carrier dynamics are also quite interesting as recently investigated by others[54-56]; such dynamics from the perspective of fs coherent scattering and imaging will be reported elsewhere.

The key results obtained are given in Fig. 10, where, to identify the lattice and CDW dynamics we selectively examine sharp scattering features employing the ultrafast crystallography modality[39]. Figure 10a gives the 6-cycle-averaged diffraction patterns where the bright Bragg peaks associated the regularly ordered TaSe$_2$ triangular lattice is clearly identifiable, denoted by $\mathbf{G_{hk}}$. Meanwhile, the CDW order registers as the weak satellites, denoted by $\mathbf{Q_j}$, can be isolated for each individual $\mathbf{G_{hk}}$. These momentum state distributions, given by the structure factors $S_{\mathbf{G}_{hk}}(\mathbf{s})$ and $S_{\mathbf{Q}_j}(\mathbf{s})$ respectively with **s** representing the scattering wave-vector, present the steady-state lattice and superlattice structures and serve as the reference point in our investigation. For ultrafast investigation, given in Fig. 1b, the changes are normalized to the ground state levels obtained at delays before the pump-probe zero-of-time. With this self-normalizing scheme, indeed multi-THz signatures of coherent phonon dynamics modulating the integrated intensity of the two structure factors can be disentangled at sub-1% level.

Most intriguingly, the non-sinusoidal waveforms, in alternating high and low intensity modulations, are observed at the low fluence level. These transient signals bear a remarkable resemblance to the optically deduced oscillating transients retrieved from MoSe$_2$, reported as originating from exciting the K-edge anti-symmetric acoustic phonon, whose modal structure is schematically reproduced in Fig. 10c[51]. In the case of MoTe$_2$, the key contribution to form the non-sinusoidal wave by the multiple-order excitation are from the fundamental wave at 4.65 THz and the harmonic at $\approx$ 9.71 THz[51]. Here, this fundamental frequency is downshifted, expected with the heavier Ta ions in the lattice[57]; nonetheless, the characteristic multiple-order excitation prevails as the most predominant channel of excitation in the multi-layer 1T-TaSe$_2$.

To better decipher the coherent signatures and the related dynamics resulted from the perturbative changes in the broken-symmetry order as well as coupling to the lattice phonon bath, we resort to properly assigning the different evolution channels to different co-factors in the structure factors. The structure factor associated with the broken-symmetry CDW order, $S_{Q_j}(\mathbf{s})$, is given in the basis of the unmodified lattice form factor:

$$F(\mathbf{s}) = \sum f_L \delta(\mathbf{r} - \mathbf{L} - u_\mathbf{L}(t)) e^{i\mathbf{s}\cdot\mathbf{r}} d\mathbf{r}, \quad (8)$$

where $\mathbf{L}$ denotes the position of the undistorted lattice[38]. The dynamical evolution shown in lattice and superlattice peaks are given by lattice displacement field $\mathbf{u}_L = \mathbf{u}_q + \mathbf{u}_\eta$, depicted by two different types of lattice waves: the lattice phonons ($\mathbf{u}_q$) or collective field modes of the broken-symmetry order ($\mathbf{u}_\eta$) with momentum wavevector $\{\mathbf{q}\}$ and $\{\mathbf{k}\}$ given by $\mathbf{q} = \mathbf{s} - \mathbf{G}_{hk}$ and $\mathbf{k} = \mathbf{s} - \mathbf{G}_{hk} - \mathbf{Q}_j$, referencing to the central Bragg reflection $\mathbf{G}_{hk}$.

In this independent mode picture, we first understand the carrier wave effects, unrelated to the broken-symmetry collective order changes. Such effects underscoring the coherent $\mathbf{u}_q(t)$ dynamics are given by the common co-factors in $S_{\mathbf{G}_{hk}}$ and $S_{\mathbf{Q}_j}$. Indeed, the appearance of the phase locking in the observed $S_{\mathbf{G}_{hk}}(\mathbf{s}, t)$ and $S_{\mathbf{Q}_j}(\mathbf{s}, t)$ reflects this. To better evaluate the lattice displacement and frequency responses, we first subtract the non-oscillating transient (in dashed line) to examine the coherent modes. The key oscillating signatures can be isolated as the superposition of two LA mode frequencies at 3.1 and 6.2 THz from a direct Fourier analysis as well as model reconstruction given in Fig. 10e in grey curve. We could determine the LA mode amplitude based on the intensity modulation[38]. The ≈ 0.5% amplitude change, as observed in Fig. 10b, gives the coherent phonon amplitude at the level of $5\times10^{-4}$ Å. Meanwhile, incoherent coupling to lattice phonon bath also presents in the $\mathbf{u}_q$-channel; here, the incoherent summing leads to the Debye-Waller factor (DWF), typically modelled in terms of 'thermal' RMS amplitude $u_T$, as an exponential decay given by momentum-dependence $e^{-|\mathbf{G}_{hk}|^2 u_T^2}$. Here, this direct coupling to the phonon bath could be judged from the sub-ps drop in the baseline levels both in $S_{\mathbf{G}_{hk}}(\mathbf{s},t)$ and $S_{\mathbf{Q}_j}(\mathbf{s},t)$. Accordingly, from following the decay at different momenta an RMS incoherent amplitude $u_T \approx 2\times10^{-4}$ Å is deduced.

Now we turn our attention to the corresponding evolution of the density wave state. We expect even in the gentle excitation regime, the direct promotion of carries condensed within the CDW collective to the excited orbitals shall weaken the strength of CDW. From the symmetry-breaking perspective, this effect, involving $\mathbf{u}_\eta$, manifests in the counteracting movements from $S_{\mathbf{Q}_j}$ (a reduction) and $S_{\mathbf{G}_{hk}}$ (an increase)[58]; see the complementary trends of baselines. The $S_{\mathbf{Q}_j}(\mathbf{s},t)$ intensity change reflects the order parameter suppression, by $3\times10^{-3}$ Å, to occur on the same sub-100 fs timescales. The rapid, collective response is an indication that the alternative displacive effect is already at play even at a low excitation level. Indeed, by nearly doubling the excitation fluence, the nonlinear quantum scattering signatures are washed out and, in its place, the broader ranges of phonon excitations are visible. We note that responses dominated by the displacive mechanism have been very recently investigated by ultrafast angle-resolved photoemission spectroscopy[54]. Upon applying IR pulse excitation just above the gap energy (0.6 eV), a transient nonequilibrium valence-band electronic manifold is shown to promptly settle in, with a transient electronic temperature up to 0.4 eV. Under this deformation stress a reduction of CDW energy gap, by an amount of up to ≈22%, occurs in ≈ 0.3 ps, followed by a short recovery to ground state on just 1 ps timescale. Meanwhile, a separate UED investigation focussed on the phase transition and excited the system with a significantly higher intensity, which is in a different regime and find charge transport across the stacking direction, establishing a 3D response associated with commensurate-to-incommensurate transition[56].

Basing off the transient dynamics, the key subject of evaluating the RF-optics-incorporated beamline performance is discussed here. First, we determine the dose to be 3 e/μm² and a coherence length of 5 nm based off the "atomic grating" approach[19, 25] adopted previously in conducting the electron bunch characterization. These give a transverse emittance $\varepsilon_x \approx 5.7$ nm, which largely is aligned with the expectation from the previous results[19]. For the temporal resolution, it is evident the RF compression has reached the sub-100 fs resolution simply judging from the initial response time in the UED signatures. Based on the visibility of the changes observed, we may retrieve the temporal resolution more quantitatively. A useful model-independent metric can be employed in which the resolution is evaluated by convoluting the data with a gaussian sigma $\sigma_c$ until the high harmonic transient signatures are washed out, which sets a bound for the response time to be $\sigma_c/\sqrt{2}$ in RMS. Through this approach, we give a temporal resolution of ≈50 fs (FWHM) based on the minimum level convolution at ≈30 fs RMS to smear out the high harmonic signatures, as given in the results plotted as the dashed line in Fig. 10e.

## VI. Summary

We have presented justification for the high-current efficiency systems and set bounds for the performance in the RF-optics-enabled UEM systems being implemented at MSU. It may seem reasonable to expect an even more stable RF circuit seeding the RF optics with better tuning of the PID control, especially within the PID-1 system, to remove the telegraph noises in our current implementation. As the system performance shall continue to improve, perhaps, a more important understanding gathered here is the correlation we have established between reducing the RF circuit instability and the performance, confirming the notation that the RF-optics-enabled UEM system can be operated in regimes near the low emittance floor offered by the laminar space-charge flow. The realization has ramification in future implementation of multimodality UEM based on the phase-space manipulation protocol as outlined here. In the case of studying coherent phonons, the successful implementation of the time-domain spectroscopy with diffraction and imaging modalities shall complement the optical studies; in particular, the direct structure-based probe can set the scale of change in the lattice degree of freedom and potentially the symmetry of the excited modal structures, e.g. see Ref. [51, 59], that has been a central topic in understanding the degree in which the momentum-dependent quantum scattering process may constitute as a viable means for coherent control of electronic processes. Continuously mapping the active coherent modes and their transformation upon increasing laser fluence *en route* to phase transition likely will provide deeper insight into the symmetry-breaking mechanisms in layered quantum materials.

## Acknowledgements

We acknowledge Florian Diekmann and Kai Rossnagel for providing samples for the microscope beam test, and discussion with Shuaishuai Sun, and Chase Boulware. Work at the Department of Physics and Astronomy at MSU are supported by the U.S. Department of Energy Basic Energy Sciences Program under grant Nos. DE-FG0206ER46309 and SC0018529; the facility support is from U.S. National Science Foundation under MRI grants DMR 1126343 and 1625181, and the MSU Foundation. Work at FRIB is supported by the U.S. Department of Energy Office of Science under Cooperative Agreement DE-SC0023633, the State of Michigan, and Michigan State University.